\documentstyle[fleqn,aps]{revtex} 
\begin{document} 
\title{Analysis of Density Matrix reconstruction in NMR Quantum Computing} 
\author{G. L. Long$^{1,2,3,4}$, H. Y. Yan$^{1,2}$ and Yang Sun$^{5,1,2}$} 
\address{ 
$^{1}$Department of Physics, Tsinghua University, Beijing 100084, P.R. China\\ 
$^{2}$Key Laboratory for Quantum Information and Measurement, 
 MOE, Beijing 100084, P.R. China\\ 
$^{3}$Institute of Theoretical Physics, Academia Sinica, Beijing 100080, P.R. China\\ 
$^{4}$Center for Atomic Molecular and Nano Sciences, Tsinghua University, Beijing 100084, 
P.R. China\\ 
$^{5}$Department of Physics and Astronomy, University of Tennessee, Knoxville,  
Tennessee 37996} 
\date{\today} 
\maketitle		 

\begin{abstract} 
Reconstruction of density matrices is important in NMR quantum computing. An analysis is 
made for a 2-qubit system by using the error matrix method. It is found that the state tomography method 
determines well the parameters that are necessary 
for reconstructing the density matrix in NMR quantum computations. 
Analysis is also made for a simplified state tomography procedure that uses fewer
read-outs. 
The result of this analysis with the error 
matrix method demonstrates that a satisfactory 
accuracy in density matrix reconstruction can be achieved even in a measurement with the number of 
read-outs being largely reduced. 
\end{abstract}

\section{Introduction} 

The study of quantum computers has attracted considerable attention since Shor 
in 1994 introduced a quantum mechanical algorithm for efficient factoring of large 
numbers \cite{r1}. 
In another remarkable work, Grover in 1996 discovered that quantum mechanics can help to  
speed up data search in an unsorted database \cite{r2,r3}. 
Quantum mechanics that, for nearly a centenary, has been a basic tool
for understanding the microscopic world 
is becoming a powerful new weapon for computation, communication 
and information-processing.

How to realize quantum computing experimentally has sparked an explosion of interest. Among 
many proposed physical systems to implement quantum computation, such as trapped 
ions, optical photons, quantum dots, and so on, 
the NMR quantum computation is particularly attractive 
because nuclear spins are extremely well isolated from their environment and readily 
manipulated with modern NMR techniques. Comprehensive algorithm realizations have been 
accomplished for the Deutsch-Jozsa algorithm in 2-qubit and 3-qubit 
systems \cite{r4,r4p1,r4p2} and a 5-qubit system \cite{r4p3}, and for Grover's algorithm  
with 2-qubit \cite{r5,r5p} and 3-qubit systems \cite{r5p2}. The order-finding problem in 
a 5-qubit system has recently been demonstrated in NMR \cite{r66}, 
and a Cat state is prepared in 
a 7-qubit system \cite{r77}. 

The fundamental elements for information-processing in NMR are two-level nuclear spins that 
are bound together in a single molecule. Put forward in 1997 by Cory {\it et al.} \cite{r6} and 
Gershenfeld {\it et al.} \cite{r7}, the NMR scheme uses bulk numbers of molecules. 
The NMR techniques cannot control the quantum states of individual 
molecules; instead, all the molecules in the sample are manipulated in parallel. 
In fact, a liquid NMR 
sample is initially in a thermal equilibrium at room temperature. The directions of the 
nuclear spins have a Boltzman distribution and are not polarized along the 
strong magnetic field. 

Reconstruction of quantum state, or quantum state tomography, is not only important in quantum computation, but also important for general purposes. It has attracted many investigations. For instance, discrete Wigner function has been used to infer the quantum states of finite-dimensional systems from measurement by Leonhardt\cite{rleonhardt1,rleonhardt2}. Walser et al have proposed a scheme to reconstruct the state of  a single quantized cavity mode\cite{rwalser}. Amiet and Weigert have studied various aspects of  reconstruction of density matrix for a spin state in pure or mixed states\cite{ramiet1,ramiet2,ramiet3}. Reconstruction of state of light has been extensively studied by Leonhardt\cite{leonhardtbook}.

In NMR quantum computation, the liquid ensemble is described by the density matrix. The state of the NMR computer can 
be obtained by state tomography technique\cite{r8}. In order to extract the density matrix, for example, for a 2-qubit system, 18 read-outs have to be performed. In general, for an $n$-qubit system, construction of the density matrix requires $3^n\times n$ read-outs.  After 
signal read-out, the area of the spectrum is integrated and the density matrix is 
reconstructed through numerical methods. Obviously, the amount of work in experiment and 
post-processing is huge when $n$ becomes moderately large.

In this paper, we study the reconstruction of density matrix in NMR. The mathematical structures of the density matrix reconstruction is analyzed. We found that the full state tomography read-outs, which consists of 18 read-outs for a 2-qubit system, 
is well conditioned.  However, a smaller number of read-outs can also determine the density matrix well. We compared the results of  2-qubit density matrix in NMR experiment with the full read-outs and a much reduced read-outs respectively and found that they agree with the theoretical analysis.

The paper is arranged as 
follows. In Section II, we describe the procedures to construct the density matrix. We analyze the  density matrix reconstruction procedure in section III. In Section IV, we discuss the reduction of number of read-outs in constructing the density matrix. We apply  the analysis to a 2-qubit NMR experiment and the results indicate that it is possible to construct a density matrix with much reduced number of read-outs and at the same the loss of accuracy is small.  
  Finally in Section V, we give a summary.

\section{Reconstruction of density matrix} 
 
In an NMR measurement, each read-out pulse can only give some off-diagonal matrix elements of 
the density matrix. 
To obtain the rest matrix elements, one has to 
rotate the original density matrix through rotational operations. 
In a 2-qubit system, in order to construct the density 
matrix, one needs to perform the following operations \cite{r8}: II, IX, IY, XI, XX, XY, YI, 
YX, and YY. 
Here, I, X and Y stand for, respectively, the identity operation, 
a 90 degree rotation about the x-axis,
and a 90 degree rotation about the y-axis. Thus, in a state tomography, 
these operations are performed before 
NMR read-out measurements. 
We explain this through an example below.  

Suppose that we use the nuclear spins of H and P in a phosphorous acid as our 
qubits. Since for a usual NMR system, only one nuclear spin can be measured at a time, we have to
perform the measurement separately for the two nuclear spins. 
We first start a computation and do a measurement on H at the required stage. 
We then restart the computation from the beginning and measure the signal 
corresponding to P. Next, we restart the computation, but this time perform the operation IX
at the required stage before measuring the signal. This process is separately carried out  
for H and P nuclear spins, and  
the nine operations are successively performed for each of them. 
This means that we have to 
perform totally $9\times 2 =18$ read-outs.  

Taking $|\uparrow\rangle=|1\rangle$ and $|\downarrow\rangle=|0\rangle$, the matrices of 
the nine operations are: 
\begin{eqnarray} 
 II=
     \left[ 
	   \begin{array}{cccc} 
	     1&0&0&0\\ 
		 0&1&0&0\\ 
		 0&0&1&0\\ 
		 0&0&0&1 
	   \end{array} 
	  \right], 
&IX=
      \left[ 
	   \begin{array}{cccc} 
	    \frac{1}{\sqrt{2}}&-\frac{i}{\sqrt{2}}&0&0\\ 
		-\frac{i}{\sqrt{2}}&\frac{1}{\sqrt{2}}&0&0\\ 
		0&0& \frac{1}{\sqrt{2}}&-\frac{i}{\sqrt{2}}\\ 
		0&0&-\frac{i}{\sqrt{2}}&\frac{1}{\sqrt{2}} 
	   \end{array} 
	   \right], 
&IY=
      \left[ 
       \begin{array}{cccc} 
	    \frac{1}{\sqrt{2}}&\frac{1}{\sqrt{2}}&0&0\\ 
		-\frac{1}{\sqrt{2}}&\frac{1}{\sqrt{2}}&0&0\\ 
		0&0&\frac{1}{\sqrt{2}}&\frac{1}{\sqrt{2}}\\ 
		0&0&-\frac{1}{\sqrt{2}}&\frac{1}{\sqrt{2}} 
	\end{array} 
	\right],\nonumber\\  
 XI=
      \left[ 
       \begin{array}{cccc} 
	 \frac{1}{\sqrt{2}}&0&-\frac{i}{\sqrt{2}}&0\\ 
	 0&\frac{1}{\sqrt{2}}&0&-\frac{i}{\sqrt{2}}\\ 
	 -\frac{i}{\sqrt{2}}&0&\frac{1}{\sqrt{2}}&0\\ 
	 0&\frac{i}{\sqrt{2}}&0&\frac{1}{\sqrt{2}} 
       \end{array} 
       \right], 
&XX=
      \left[ 
       \begin{array}{rrrr} 
	   \frac{1}{2}&-\frac{i}{2}&-\frac{i}{2}&-\frac{1}{2}\\ 
	   -\frac{i}{2}&\frac{1}{2}&-\frac{1}{2}&-\frac{i}{2}\\ 
	   -\frac{i}{2}&\frac{1}{2}&\frac{1}{2}&-\frac{i}{2}\\ 
	   -\frac{1}{2}&-\frac{i}{2}&-\frac{i}{2}&\frac{1}{2} 
	   \end{array} 
	   \right], 
&XY= 
      \left[ 
       \begin{array}{rrrr} 
	   \frac{1}{2}&\frac{1}{2}&-\frac{i}{2}&\frac{i}{2}\\ 
	   -\frac{1}{2}&\frac{1}{2}&\frac{i}{2}&-\frac{i}{2}\\ 
	   -\frac{i}{2}&-\frac{i}{2}&\frac{1}{2}&\frac{1}{2}\\ 
	   \frac{i}{2}&-\frac{i}{2}&-\frac{1}{2}&\frac{1}{2} 
	  \end{array} 
	  \right],\\ 
 YI= 
      \left[ 
       \begin{array}{cccc} 
	   \frac{1}{\sqrt{2}}&0&\frac{1}{\sqrt{2}}&0\\ 
	   0&\frac{1}{\sqrt{2}}&0&\frac{1}{\sqrt{2}}\\ 
	   -\frac{1}{\sqrt{2}}&0&\frac{1}{\sqrt{2}}&0\\ 
	   0&-\frac{1}{\sqrt{2}}&0&\frac{1}{\sqrt{2}} 
	   \end{array} 
	   \right], 
&YX= 
      \left[ 
       \begin{array}{rrrr} 
	   \frac{1}{2}&-\frac{i}{2}&\frac{1}{2}&-\frac{i}{2}\\ 
	   -\frac{i}{2}&\frac{1}{2}&-\frac{i}{2}&\frac{1}{2}\\ 
	   -\frac{1}{2}&\frac{i}{2}&\frac{1}{2}&-\frac{i}{2}\\ 
	   \frac{i}{2}&-\frac{1}{2}&-\frac{i}{2}&\frac{1}{2} 
	   \end{array} 
	   \right], 
&YY= 
      \left[ 
       \begin{array}{rrrr} 
	   \frac{1}{2}&\frac{1}{2}&\frac{1}{2}&\frac{1}{2}\\ 
	   -\frac{1}{2}&\frac{1}{2}&-\frac{1}{2}&\frac{1}{2}\\ 
	   -\frac{1}{2}&-\frac{1}{2}&\frac{1}{2}&\frac{1}{2}\\ 
	   \frac{1}{2}&-\frac{1}{2}&-\frac{1}{2}&\frac{1}{2} 
	   \end{array} 
	   \right].\nonumber 
\end{eqnarray} 
	    
Let us assume that the density matrix of the system takes the form  
\begin{equation} 
\rho=\left[ 
        \begin{array}{rrrr} 
		 \rho_{11}& \rho_{12}&\rho_{13}&\rho_{14}\\ 
		 \rho_{12}^{\ast}&\rho_{22}&\rho_{23}&\rho_{24}\\ 
		 \rho_{13}^{\ast}&\rho_{23}^{\ast}&\rho_{33}&\rho_{34}\\ 
		 \rho_{14}^{\ast}&\rho_{24}^{\ast}&\rho_{34}^{\ast}&\rho_{44} 
		\end{array} 
		\right]= 
	\left[ 
        \begin{array}{cccc} 
		 x_{1}&x_{2}+ix_{11}&x_{3}+ix_{12}&x_{4}+ix_{13}\\ 
		 x_{2}-ix_{11}&x_{5}&x_{6}+ix_{14}&x_{7}+ix_{15}\\ 
		 x_{3}-ix_{12}&x_{6}-ix_{14}&x_{8}&x_{9}+ix_{16}\\ 
		 x_{4}-ix_{13}&x_{7}-ix_{15}&x_{9}-ix_{16}&x_{10} 
		\end{array} 
		\right]. 
\label{density}
\end{equation} 
The NMR read-out signal can only give $x_{2}+ix_{11}$ and $x_{9}+ix_{16} $ for the nuclear spin 
of P, and $x_{3}+ix_{12}$ and $x_{7}+ix_{15}$ for the nuclear spin of H.  
For example, the element $\rho_{13}$ in Eq. (\ref{density}) 
corresponds to the left peak of the spectrum of H, 
while the element $\rho_{24}$ to the right peak of it. Similarly, the element 
$\rho_{12}$ corresponds to the left peak of the spectrum of P, and the element $\rho_{34}$ 
to the right peak of the same spectrum. To obtain other 
elements in the density matrix, we have to perform one of the nine operations for the system 
so that the desired elements are transformed to the positions labeled as 12, 13, 24 and 34 
in the density matrix, and thus can be measured. 
The read-out gives the elements $\rho_{12}'$, $\rho_{13}'$ and $\rho_{24}'$, 
$\rho_{34}'$ in the rotated density matrix. These rotated matrix elements 
are linear combinations of the original matrix elements. 
For each of the nine operations, one makes two measurements for the 
nuclear spins of H and P, and each measurement provides two matrix elements which contains a real 
and an imaginary part. Altogether, we finally obtain $4\times 9\times 2=72$ equations with 16 
unknowns. 

The coefficients in these equations form a matrix with the size of $72\times 16$. 
The rank of this coefficient matrix is only 15 since the matrix can be 
added with a matrix of a constant times the unit matrix without changing the results. In 
practice, we can add one more equation by letting the trace of the density matrix be 
1 after normalizing the integrations of the spectrum. Thus, we have totally 73 
equations. To determine the elements in the density matrix, one needs to solve the 
following set of linear equations for $x_{i}$:  
\begin{equation} 
\sum _{i=1}^{16}A_{\alpha i}x_{i}=B_{\alpha},  \qquad \{\alpha=1, 2, \ldots 73\}, 
\label{b} 
\end{equation} 
where $A_{\alpha i}$ is the coefficient of $x_i$ in the $\alpha$-th equation and 
it varies with read-out for different rotations, and 
$B_{\alpha}$ is the integrated area of the spectrum. 
There are certainly redundant expressions in (\ref{b}) since the 
number of equations is more than the number of unknowns. The standard way of dealing 
with this problem is to use the least square fitting procedure 
that is widely used in various problems in science and engineering.
We minimize the quantity $\chi^{2}$ defined as 
\begin{equation} 
\chi^{2}=\sum _{\alpha}(\sum _{i=1}^{16}A_{\alpha i}x_{i}-B_{\alpha})^{2}. 
\end{equation} 
To find the minimum, we carry out a variation procedure on $\chi^{2}$ with respect to
all parameters, which gives 
\begin{equation} 
\sum _{j=1}^{16}C_{ij}x_{j}=b_{i}, 
\label{newb}
\end{equation} 
where  
\begin{equation} 
C_{ij}=\sum _{\alpha}A_{\alpha i}A_{\alpha j}, \qquad  
{\rm{and}} \quad  
b_{i}=\sum _{\alpha}B_{\alpha}A_{\alpha i}. 
\label{c}
\end{equation} 
The number of equations in (\ref{newb}) is now equal to the number of unknowns. These linear
equations can be 
solved by standard numerical method such as the Gaussian elimination. 
In principle, such problems can be solved in this way.
However, in most of our cases, not every parameter can be well determined in experiment.
In the least square fitting procedure, these less well-determined matrix elements might 
possibly bring spurious values into the numerical calculation while  
leaving the $\chi^2$ value small. In other words, the $\chi^2$ might not be 
very sensitive in response to a big variation in experiment.  For instance, if we have a set of equations, x+y=e1, x+1.001y=e2, this set of equations determines the sum of x and y very well, and the difference between x and y is poorly determined. The $\chi^2$ is insensitive to a change in x-y while leaving x+y unchanged.

\section{An analysis of the density matrix reconstruction} 
 
In order to obtain the density matrix, one needs to use the least-square fitting method\cite{r8p} to derive them from the experimental data. Similar problem occurs actually in other fields of science as well. Here, 
we adopt a variant of the least square-fitting method from nuclear physics to solve the problem. 
In the large-scale nuclear shell model calculations, 
Wildenthal {\it et al.} employed the error matrix 
method \cite{r7p} to analyze sensitivity of the nuclear structure to experimental input. 
In their method, instead of solving Eq. (\ref{b}) 
directly, one first solves the eigenvalue problem for $C$ defined in (\ref{c}), 
\begin{equation} 
UCU^{+}=C_{d}. 
\end{equation} 
Here, $U$ is the unitary matrix that diagonalizes $C$. $C_{d}$ is called the error
matrix \cite{r7p}. Then Eq. (\ref{b}) 
becomes  
\begin{eqnarray} 
C_{d}y=b^{'}, 
\end{eqnarray}  
where $y_{i}=U_{ij}x_{j}$ and $b^{'}_{i}=U_{ij}b_{j}$. $b^{'}_{i}$ contains experimental 
information.   Since $C_{d}$ is diagonal, we can determine $y_{i}$ by 
\begin{eqnarray} 
y_{i}=\frac{b^{'}_{i}}{(C_{d})_{ii}}. 
\end{eqnarray} 
$b^{'}_{i}$ contains experimental uncertainties, and any change in $b_{i}^{'}$ will cause $y_{i}$ 
to change. However, if the diagonal matrix element $(C_{d})_{ii}$ is large (say, of the order of 
1), $y_{i}$ will be insensitive to changes in $b_{i}^{'}$, and thus it can be well determined. 
Conversely, a small $(C_{D})_{ii}$ (say, 0.001) means that the corresponding $y_{i}$ is 
very sensitive to experiment, and any small variation in $b'$ will cause a big change in 
$y$. In this case, $y$ is not well determined, and special effort is necessary 
to ensure that $b'$ is sufficiently accurate. If this is not possible, then 
during the fitting process, the $y'$s that corresponds to small eigenvalues of $C$ will be
kept constant by physical considerations. In nuclear structure studies, the 
critical constant value was chosen as 0.001. 
 
The $C$ matrix for a 2-qubit system with all the nine transformations performed (with 18 read-outs, 
and 73 linear equations) is 
\begin{equation}\label{a} C= \left[ 
     \begin{array}{cccccccccccccccc} 
	 3&0&0&0&\frac{1}{2}&0&0&\frac{1}{2}&0&0&0&0&0&0&0&0\\ 
	 0&5&0&0&0&0&0&0&1&0&0&0&0&0&0&0\\ 
	 0&0&5&0&0&0&1&0&0&0&0&0&0&0&0&0\\ 
	 0&0&0&4&0&0&0&0&0&0&0&0&0&0&0&0\\ 
	 \frac{1}{2}&0&0&0&3&0&0&0&0&\frac{1}{2}&0&0&0&0&0&0\\ 
	  
	 0&0&0&0&0&4&0&0&0&0&0&0&0&0&0&0\\ 
	 0&0&1&0&0&0&5&0&0&0&0&0&0&0&0&0\\ 
	 \frac{1}{2}&0&0&0&0&0&0&3&0&\frac{1}{2}&0&0&0&0&0&0\\ 
	 0&1&0&0&0&0&0&0&5&0&0&0&0&0&0&0\\ 
	 0&0&0&0&\frac{1}{2}&0&0&\frac{1}{2}&0&3&0&0&0&0&0&0\\ 
	  
	 0&0&0&0&0&0&0&0&0&0&5&0&0&0&0&1\\ 
	 0&0&0&0&0&0&0&0&0&0&0&5&0&0&1&0\\ 
	 0&0&0&0&0&0&0&0&0&0&0&0&4&0&0&0\\ 
	 0&0&0&0&0&0&0&0&0&0&0&0&0&4&0&0\\ 
	 0&0&0&0&0&0&0&0&0&0&0&1&0&0&5&0\\ 
	  
	 0&0&0&0&0&0&0&0&0&0&1&0&0&0&0&5\\ 
	 \end{array} 
	 \right]. 
\end{equation}	  
Solving the eigenvalue problem, and we obtain the 16 eigenvalues of $C$. They are: 4, 4, 4, 4, 4,  
6, 6, 3, 3, 2, 6, 4, 4, 4, 4, 6. 
The $y$'s are the combinations of the $x$ coefficients 
\begin{eqnarray} 
\label{d} 
y_1&=&-0.37x_{3}+0.82x_{4}-0.24x_{6};\nonumber\\ 
y_2&=&-0.19x_{1}+0.41x_{2}-0.17x_{3}-0.32x_{4}-0.19x_{5} 
-0.59x_{6}+0.17x_{7}-0.19x_{8}-0.41x_{9}-0.19x_{10};\nonumber\\ 
y_3&=&-0.0087x_{1}-0.093x_{2}-0.57x_{3}-0.39x_{4}-0.0087x_{5} 
+0.42x_{6}+0.57x_{7}-0.0087x_{8}+0.093x_{9}-0.0087x_{10};\nonumber\\ 
y_4&=&0.30x_{1}-0.30x_{2}-0.093x_{3}-0.26x_{4}+0.30x_{5} 
-0.61x_{6}+0.093x_{7}+0.30x_{8}+0.30x_{9}+0.30x_{10};\nonumber\\ 
y_5&=&-0.35x_{1}-0.48x_{2}+0.027x_{3}-0.035x_{4}-0.35x_{5} 
-0.20x_{6}-0.027x_{7}-0.35x_{8}+0.48x_{9}-0.35x_{10};\nonumber\\ 
y_6&=&-0.55x_{2}+0.44x_{3}+0.44x_{7}-0.55x_{9};\nonumber\\ 
y_7&=&-0.44x_{2}-0.55x_{3}-0.55x_{7}-0.44x_{9};\nonumber\\ 
y_8&=&0.71x_{5}-0.71x_{8};\nonumber\\ 
y_9&=&-0.71x_{1}+0.71x_{10};\nonumber\\ 
y_{10}&=&0.50x_{1}-0.50x_{5}-0.50x_{8}+0.50x_{10};\nonumber\\ 
y_{11}&=&0.71x_{12}+0.71x_{15};\nonumber\\ 
y_{12}&=&-0.71x{12}+0.71x_{15};\nonumber\\ 
y_{13}&=&x_{13};\nonumber\\ 
y_{14}&=&x_{14};\nonumber\\ 
y_{15}&=&0.71x_{11}-0.71x_{16};\nonumber\\ 
y_{16}&=&0.71x_{11}+0.71x_{16}.\nonumber 
\end{eqnarray} 
Some of the $y'$s are directly the $x$ parameters in the density matrix such as $y_{13}$ 
and $y_{14}$, and others are the combinations of the $x$ parameters. They are all 
well determined by the experimental data.

\section{reduction of experimental read-outs}

In the process of the density matrix reconstruction discussed above, 
each signal read-out provides us two equations. By performing 
all the nine operations for P and H, and plus the 
normalization condition, we have 73 equations. 
However, these equations are over-determined. The question is if it is possible 
to determine the density matrix with less read-outs. It will be very interesting to see what is 
the minimum number of operations to determine the density matrix without loss of much 
accuracy. Intuitively, we may think about 4 read-outs because there are altogether 16 unknowns. 
However, a detailed analysis of the rank of equations indicated that any 4 
read-outs combined together can not provide sufficient numbers of independent equations. 
The minimum number of read-outs to be performed is 5. In table (\ref{t1}) we give the various  set of 5 read-outs. In the table, 
operation 1 is the II with H signal acquisition(see table caption for details).
From the table we see that not arbitrary combination of 5 operations are possible.

 To see how well the density matrix is determined by a smaller number of read-outs, 
we choose a  6 read-outs case. The six read-outs are:  II, IX, IY, XX for one of the nuclear spins (H), and II, IX for
the other (P). In this situation, the corresponding $C$ matrix becomes: 
\begin{equation} 
C=\left[ 
    \begin{array}{cccccccccccccccc} 
	\frac{11}{8}&0&0&0&\frac{7}{8}&0&0&\frac{7}{8}&0&\frac{7}{8}&0&0&0&0&0&0\\ 
	0&2&0&0&0&0&0&0&0&0&0&0&0&0&0&0\\ 
	0&0&\frac{5}{2}&0&0&0&\frac{3}{2}&0&0&0&0&0&0&0&0&0\\ 
	0&0&0&1&0&0&0&0&0&0&0&0&0&0&0&0\\ 
	\frac{7}{8}&0&0&0&\frac{11}{8}&0&0&\frac{7}{8}&0&\frac{7}{8}&0&0&0&0&0&0\\ 
	 
	0&0&0&0&0&1&0&0&0&0&0&0&0&0&0&0\\ 
	0&0&\frac{3}{2}&0&0&0&\frac{5}{2}&0&0&0&0&0&0&0&0&0\\ 
	\frac{7}{8}&0&0&0&\frac{7}{8}&0&0&\frac{3}{8}&0&\frac{7}{8}&0&0&0&0&0&0\\ 
	0&0&0&0&0&0&0&0&2&0&0&0&0&0&0&0\\ 
	\frac{7}{8}&0&0&0&\frac{7}{8}&0&0&\frac{7}{8}&0&\frac{11}{8}&0&0&0&0&0&0\\ 
	 
	0&0&0&0&0&0&0&0&0&0&\frac{3}{2}&0&0&0&0&-\frac{1}{2}\\ 
	0&0&0&0&0&0&0&0&0&0&0&2&0&0&1&0\\ 
	0&0&0&0&0&0&0&0&0&0&0&0&\frac{3}{2}&-\frac{1}{2}&0&0\\ 
	0&0&0&0&0&0&0&0&0&0&0&0&-\frac{1}{2}&\frac{3}{2}&0&0\\ 
	0&0&0&0&0&0&0&0&0&0&0&1&0&0&2&0\\ 
	0&0&0&0&0&0&0&0&0&0&-\frac{1}{2}&0&0&0&0&\frac{3}{2} 
	\end{array} 
	\right]. 
\end{equation} 
The eigenvalues of $C$ matrix are: 1, 2, 1, $\frac{1}{2}$, $\frac{1}{2}$, 1, 4, 2, 4, 
$\frac{1}{2}$, 1, 3, 1, 2, 1, 2. 
It can be seen that the eigenvalues are all quite big.
The $y$ expressions corresponding to the eigenvalues are: 
\begin{eqnarray} 
y_1&=&-x_{6};\nonumber\\ 
y_2&=&x_{2};\nonumber\\ 
y_3&=&x_{4};\nonumber\\ 
y_4&=& -0.79x_{1}+0.21x_{5}+0.58x_{8};\nonumber\\ 
y_5&=& -0.21x_{1}+0.79x_{5}-0.58 
 x_{8};\nonumber\\ 
y_6&=& 0.71x_{3}-0.71x_{7};\nonumber\\ 
y_7&=& 0.71x_{3}+0.71x_{7};\nonumber\\ 
y_8&=&x_{9};\nonumber\\ 
y_9&=& -0.50x_{1}-0.50x_{5}-0.50x_{8}-0.50x_{10};\nonumber\\ 
y_{10}&=& -0.29x_{1}-0.29x_{5}-0.29x_{8}+0.87x_{10};\nonumber\\ 
y_{11}&=& -0.71x_{12}+0.71x_{15};\nonumber\\ 
y_{12}&=&  0.71x_{12}+0.71x_{15};\nonumber\\ 
y_{13}&=& -0.71x_{13}-0.71x_{14};\nonumber\\  
y_{14}&=& -0.71x_{13}+0.71x_{14};\nonumber\\ 
y_{15}&=& -0.71x_{11}-0.71x_{16};\nonumber\\ 
y_{16}&=& -0.71x_{11}+0.71x_{16}.     
\end{eqnarray} 
The density matrix elements are as 
well determined as the one with 18 read-outs. However, the saving in the numbers of read-outs is 
great ($18-6=12$).  Because we have less read-outs now, statistics in this case 
is surely poorer, and this will reduce 
accuracy of the parameters determined. Nevertheless, because of the intrinsic mathematical 
structure, the uncertainties in the experimental data affect the parameters 
insensitively, and the loss in accuracy is not very big.

As an example, we have analyzed the density matrix for a 2-qubit system 
in an NMR experiment \cite{r10}. The theoretical prediction of the density matrix is 
\begin{equation}\rho _{\rm th}=\left[ 
            \begin{array}{rrrr} 
			0.31&0.31&0.31&-0.063-0.13i\\ 
			0.31&0.31&0.31&-0.063-0.13i\\ 
			0.31&0.31&0.31&-0.063-0.13i\\ 
			-0.063+0.13i&-0. 063+0.13i&-0.063+0.13i&0.063 
			\end{array} 
			\right]. 
\end{equation} 
When constructing the density matrix using all 73 equations, we obtain 
\begin{equation}\rho _{\rm all}=\left[ 
            \begin{array}{rrrr} 
			0.36&0.33-0.087 i&0.31+0.037 i&-0.034-0.22i\\ 
			0.33+0.087i&0.30&0.28-0.022i&-0.079-0.14i\\ 
			0.31-0.037i&0.28+0.022i &0.23&-0.044-0.13i\\ 
			-0.034+0.22i&-0. 079+0.14i&-0.044+0.13i&0.12 
			\end{array} 
			\right]. 
\end{equation} 
If we take only 49 equations out of the 73 equations, which means that we take only 12  
read-outs (which contains all 9 read-outs for H and the II, IX, IY read-outs for P) 
instead of the complete 18 read-outs, the density matrix is 
\begin{equation}\rho _{12}=\left[ 
            \begin{array}{rrrr} 
			0.37&0.31-0.13i&0.29+0.034i&-0.074-0.17i\\ 
			0.31+0.13i&0.28&0.37+0.0091i&-0.059-0.14i\\ 
			0.29-0.034i&0.37-0.0091i&0.25&-0.058-0.17i\\ 
			-0.074+0.17i&-0.059+0.14i&-0.058+0.17i&0.094 
			\end{array} 
			\right]. 
\end{equation} 
We see that the experimental density matrices are nearly identical for the two cases. 
Defining the error measurement  
\begin{equation}
\delta =\left| \frac{\|\rho _{\rm exp}-\rho _{\rm th}\|_{2}}{\|\rho _{\rm exp}\|_{2 
}}\right|,  
\end{equation}
where $\|\|$ is the norm of a matrix, we get the error of $\rho _{\rm all}$ relative to $\rho _{\rm th}$ $ 17\%$, and  
the same error of $\rho _{12}$ to $\rho _{\rm th}$, which roughly says that $\rho  
_{\rm all}$ or $\rho_{12}$ is about $83\%$ of $\rho _{\rm th}$. Thus we have seen 
that the number of read-outs 
is reduced quite a lot, but the accuracy in the density matrix is not much influenced. 
For the extreme case with 6 read-outs only, the density matrix is: 
\begin{equation}\rho _{6}=\left[ 
            \begin{array}{rrrr} 
			0.52&0.29-0.21i&0.29+0.068i&-0.025-0.14i\\ 
			0.29+0.21i&0.16&0.37+0.085i&-0.045-0.13i\\ 
			0.29-0.068i&0.37-0.085i&0.32&0.0039-0.15i\\ 
			0.025+0.14i&-0.045+0.13i&0.0039+0.15i&0.18 
			\end{array} 
			\right]. 
\end{equation} 
Still, it is quite close to $\rho_{\rm th}$. The relative error of $\rho_{6}$ to $\rho_{\rm th}$ is $32\%$. 
We see that, although the number of equations is reduced more than a half, the errors do not increase  
as much as one would think. Since the reconstruction of the density matrix in NMR quantum 
computing is so tedious, the present work suggests a way 
to reduce the number of read-outs if the accuracy is 
not so highly required.  

\section{Summary}
In conclusion, we have analyzed the error matrix for the density matrix reconstruction in the 2-qubit NMR 
quantum computing. We have found that the number of read-outs can be reduced greatly without 
significant loss in the accuracy. Our analysis can be easily extended to NMR systems with 
a larger number of qubits. 
 
The authors are grateful for financial support from the China National Natural Science 
Foundation, the Major State Basic Research Development Program under contract no. G200077407, 
the Hangtian Science Foundation, and the Fok Ying Tung Science Foundation.

\begin{table}
\begin{center}
\caption{Minimum set of read-outs combinations. Here 1=II, 2=IX, 3=IY, 4=XI, 5=XX, 6=XY, 7=YI, 8=YX, 9=YY with H signal acquisition, and 10=II, 11=IX, 12=IY, 13=XI, 14=XX, 15=XY, 16=YI, 17=YX, 18=YY with P signal acquisition.}
\label{t1}
\begin{tabular}{ccccc|ccccc|ccccc}\hline
1 & 2 & 6 & 12 & 13 & 1 & 2 & 6 & 12 & 14 & 1 & 2 & 9 & 12 & 16\\
1 & 2 & 9 & 12 & 17 & 1 & 3 & 5 & 11 & 13 & 1 & 3 & 5 & 11 & 15 \\
1 & 3 & 8 & 11 & 16 & 1 & 3 & 8 & 11 & 18 & 1 & 5 & 6 & 11 & 13\\
1 & 5 & 6 & 12 & 13 & 1 & 5 & 11& 13 & 16 & 1 & 5 & 11& 13 & 17\\
1 & 6 & 12& 13 & 16 & 1 & 6 & 12& 13 & 18 & 1 & 8 & 9 & 11 & 16\\
1 & 8 & 9 & 12 & 16 & 1 & 8 & 11& 13 & 16 & 1 & 8 & 11& 14 & 16\\
1 & 9 & 12& 13 & 16 & 1 & 9 & 12& 15 & 16 & 2 & 3 & 4 & 10 & 14\\
2 & 3 & 4 & 10 & 15 & 2 & 3 & 7 & 10 & 17 & 2 & 3 & 7 & 10 & 18\\
2 & 4 & 6 & 10 & 14 & 2 & 4 & 6 & 12 & 14 & 2 & 4 &10 & 14 & 16\\
2 & 4 &10 & 14 & 17 & 2 & 6 & 12& 14 & 17 & 2 & 6 &12 & 14 & 18\\
2 & 7 & 9 & 10 & 17 & 2 & 7 & 9 & 12 & 17 & 2 & 7 &10 & 13 & 17\\
2 & 7 &10 & 14 & 17 & 2 & 9 & 12& 14 & 17 & 2 & 9 &12 & 15 & 17\\
3 & 4 & 5 & 10 & 15 & 3 & 4 & 5 & 11 & 15 & 3 & 4 &10 & 15 & 16\\
3 & 4 &10 & 15 & 18 & 3 & 5 &11 & 15 & 17 & 3 & 5 &11 & 15 & 18\\
3 & 7 & 8 & 10 & 18 & 3 & 7 & 8 & 11 & 18 & 3 & 7 & 10& 13 & 18\\
3 & 7 &10 & 15 & 18 & 3 & 8 & 11& 14 & 18 & 3 & 8 & 11& 15 & 18\\
4 & 5 & 9 & 15 & 16 & 4 & 5 & 9 & 15 & 17 & 4 & 6 & 8 & 14 & 16\\
4 & 6 & 8 & 14 & 18 & 4 & 8 & 9 & 14 & 16 & 4 & 8 & 9 & 15 & 16 \\
4 & 8 & 10& 14 & 16 & 4 & 8 & 11& 14 & 16 & 4 & 9 & 10 & 15& 16\\
4 & 9 & 12& 15 & 16 & 5 & 6 & 7 & 13 & 17 & 5 & 6 & 7 & 13 & 18\\
5 & 7 & 9 & 13 & 17 & 5 & 7 & 9 & 15 & 17 & 5 & 7 & 10& 13 & 17\\
5 & 7 & 11 &13 & 17 & 5 & 9 & 11& 15 & 17 & 5 & 9 & 12& 15 & 17\\
6 & 7 & 8 & 13 & 18 & 6 & 7 & 8 & 14 & 18 & 6 & 7 & 10& 13 & 18\\
6 & 7 & 12& 13 & 18 & 6 & 8 & 11& 14 & 18 & 6 & 8 & 12 & 14 & 18\\
\hline
\end{tabular}
\end{center}
\end{table}


\begin{thebibliography}{99} 
\bibitem{r1} P.W. Shor, in Proceedings of the 35th Annual Symposium on Foundations of 
 Computer Science (IEEE Computer Society Press, Los Alamitos, CA, 1994) p. 124, 
 quant-ph/9508027 12 A 
\bibitem{r2} L.K. Grover, Phys. Rev. Lett 79, 325 (1997) 
\bibitem{r3} L.K. Grover, Phys. Rev. Lett 80, 4329 (1998) 
\bibitem{r4} I.L. Chuang, L.M.K. Vandersypen, X. Zhou, D.W. Leung and S. Lloyd, 
 Nature 393, 143 (1998)  
\bibitem{r4p1} N. Linden, H. Barjat and R. Freeman, Chem. Phys. Lett. 296, 61 (1998) 
\bibitem{r4p2} K. Dorai, Arvind and A. Kumar, Phys. Rev. A 6104, 2306 (2000) 
\bibitem{r4p3} R. Marx, A.F. Fahmy, J.M. Myers, W. Bermel and S.J. Glaser, PHYS REV A 62(2000)012310 
\bibitem{r5} I.L. Chuang, N. Gershenfeld and M. Kubinec, Phys. Rev. Lett 80, 3408 (1998) 
\bibitem{r5p} J.A. Jones, M. Mosca and R. H.  Hansen, Nature 393, 344 (1998)    
\bibitem{r5p2} L.M.K. Vandersypen {\it et al.}, Appl. Phys. Lett. 76, 646 (2000) 
\bibitem{r66} L.M.K. Vandersypen, M. Steffen, G. Breyta, C.S. Yannoni, R. Cleve and I.L. Chuang, Phys. Rev. Lett. 85, 25, 5452-5455 (18 Dec 2000). Also in quant-ph/0007017 
\bibitem{r77} E. Knill, R. Laflamme, R. Martinez and C.-H. Tseng, quant-ph/9908051 
\bibitem{r6} D. Cory, A. Fahmy and T. Havel, Proc. Nat. Acad. Sci. U.S.A. 94, 1634 (1997) 
\bibitem{r7} N. Gershenfeld and I.L. Chuang, Science 275, 350 (1997)
\bibitem{rleonhardt1} U. Leonhardt, Phys. Rev. Lett., 74 (1995) 4101
\bibitem{rleonhardt2} U. Leonhardt, Phys. Rev., A53 (1996) 2998.
\bibitem{rwalser} R. Walser, J. I. Cirac and P. Zoller, Phys. Rev. lett., 77 (1996) 2658
\bibitem{ramiet1} J. P. Amiet and S. Weigert, J. Phys. A 31 (1998) L543
\bibitem{ramiet2} J. P. Amiet and S. Weigert, J. Phys. A 32 (1999) 2777
\bibitem{ramiet3} J. P. Amiet and S. Weigert, J. Phys. A 32 (1999) L269
\bibitem{leonhardtbook} U. Leonhardt, Measuring the quantum state of light (Cambridge University Press, Cambridge, 1997)
\bibitem{r8} I.L. Chuang, N. Gershenfeld, M. Kubinec and D. Leung, 
 Proc. R. Soc. Lond A 454, 447 (1998) 
 \bibitem{r8p}W. H. Press et. al., {\it Numerical recipes}, Cambridge University Press,1989
 \bibitem{r7p} W. Chung, Ph.D. thesis, Michigan State University (1976);  X. Ji and B.H. 
 Wildenthal, Phys. Rev. C37, 1256 (1988) 
\bibitem{r9} R.R. Ernst, G. Bodenhausen and A. Wokaun, {\it Principles of Nuclear Magnetic 
 Resonance in One and Two 
 Dimensions} (Oxford University Press, Oxford, 1994) 
\bibitem{r10} G.L. Long, H.Y. Yan {\it et al.}, {\it Experimental NMR realization of a generalized quantum search algorithm}, Phys. Lett. A286 (2001) 121. Also as quant-ph/0009059. 
\end{thebibliography}
\end{document}